# Derivation of the nonlinear equations for surface of fluid adhering to a moving plate withdrawn from a liquid pool

Ivan V. Kazachkov[1,2]

[1]Dept of Information Technologies and Data Analysis,
Nizhyn Gogol State University, UKRAINE, http://www.ndu.edu.ua
[2]Dept of Energy Technology, Royal Institute of Technology, Stockholm, 10044, SWEDEN,
Ivan.Kazachkov@energy.kth.se, http://www.kth.se/itm/inst?l=en_UK

## Abstract

Many technological processes include preparing some special materials' adhering to a product surface. For example, this problem is important for the magnetic tape producing, wire adhering, etc. For a surface withdrawn from the molten metal or the other liquid material there is a problem to determine a profile of a film surface. It is subject of this paper. We developed the mathematical model for the simulation of the adhering process of viscous liquid film to a slowly moving plate, which is vertically withdrawn from the molten metal or the other fluid capacity. The Navier-Stokes equations for a film flow on a surface of the withdrawn plate are considered with the corresponding boundary conditions, and the polynomial approximation is used for the film flow profile. The equations after integration across the layer of a film flow result in the system of partial differential equations for the wavy surface $\zeta(t,x)$ of a film flow, of flow rate $q(t,x)$ and of flow energy $Q(t,x)$.The derived equations are used for analysis of the nonlinear film flow that determines the quality of a fluid adhering on a surface of the withdrawn plate.

## 1. Introduction

The problem of coating the surface of a material with a thin film of a liquid has been considered for a long time in various practical problems, and then it has also gained interest theoretically. This is due to the need to apply uniform coatings of a given thickness, to estimate the entrainment of a liquid from the pool after removing the object from it, to determine an amount of a liquid on the walls of the container after pouring out the liquid from it, and many other practically important tasks. Also, many technological processes are doing some special materials' adhering to a product's surface. For example, this problem is important for the magnetic tape's producing, wire adhering, etc. For the surface withdrawn from the molten metal or from the other liquid material there is a problem to determine the film surface profile as much as possible precisely and to control it. This is a subject of the present paper.

The theory of the free coating of a Newtonian liquid on a plate was developed on the basis of a scale analysis of the flow [1] with an analysis isolated on the flow in the apical part of the meniscus where the film is captured and also the bulk liquid is transported from the depths of the basin to the surface. The film thickness and the characteristic curvature of the meniscus were expressed in terms of the capillary number $Ca=\mu u_0/\sigma$(viscous forces to capillary forces) and the dimensionless parameter $Po=\mu(g/\rho\sigma^3)^{1/4}$, where the first is the dynamic criterion and the other is the kinematic criterion (it is determined only by physical properties: gravitational, viscous and capillary).Here $\mu$, $\rho$,$\sigma$ are the dynamic viscosity coefficient, density, surface tension coefficient of the liquid, respectively, $g$ is acceleration due to gravity; $u_0$ is the characteristic velocity of a fluid flow. The authors [1] used two adjustable constants, determined by least-squares fitting with the experimental data [2], which were surprisingly "universal" for free-coating on a plate. For a range of $Ca$ the film thickness was scaled to $d_0=(\mu u_0/\rho g)^{1/2}$ asymptotes independent of $Po$. But in the small-$Ca$ limit the classical Landau-Levich law [3] is duly recovered. The roles of capillary, inertial, and gravity forces in the various regimes are playing depending on their ratio changing the regimes. The theoretical and experimental results are correlated well spanning over three orders of magnitudes both of $Ca$ and $Po$. It is interesting to note that a non-monotonous behaviour of the characteristic meniscus curvature scaled to the reciprocal film thickness, with a growth followed by a drop as a function of *Ca*, is predicted, in qualitative accordance with earlier experimental observations and computational results.

## 2. Accomplishments and challenges in the film coatings

Derjaguin [4] proposed a "load" *h*, i.e. the thickness of the film for the liquid adhering to the plate $h=(\mu u_0/\rho g\sin\alpha)^{1/2}$ ($h=d_0(\sin\alpha)^{1/2}$), assuming that the effects of inertia and surface tension are weak. In [2], an infinite plate (at an angle α to the horizontal at a constant velocity $u_0$) is considered from an infinite pool of a viscous liquid, where the above formula obtained from the Stokes equations within the boundaries of small slopes of the plate (without this assumption, the formula is invalid). The problem was shown to have infinitely many stable solutions; all of them are stable but only one corresponds to the above formula. This particular stable solution can be distinguished only by comparing it with a self-similar solution describing the non-steady part of the film flow between the pool and the tip of the film. Although the area of the near-pool region in which the stable state is established expands with time, the upper non-steady part of the film (its thickness decreases to the tip) expands faster as it was shown. It occupies most of the plate; therefore an average thickness of the film is 1.5 times smaller than the load.

For the case of thick films, the formula [4] has been given without strict derivation, showing, in particular, that in this case the thickness of the layer is independent of a surface tension of a liquid. In [2]it was derived more in detail considered the profile of a liquid layer which remains on the wall of a vessel, inclined at an angle to the horizon, at a time *t* after the level of the liquid has begun to recede. It was supposed that the condition $dh/dx<<1$for a thickness *h* of a film at the given point, is satisfied everywhere, except at the place, where the film goes over into the free volume of a liquid. Publication [4] was delayed due to the discovery of divergences from experiment, the explanation for which was found later. The experimental data [2] fully confirm the theory including the numerical coefficients.

The work of Landau and Levich [3] (1942) initiated the fundamental theoretical, as well as experimental investigation of a flow of the thin liquid film entrained by a steady withdrawal of a flat plate from a liquid bath. The existing theories are based on a linearization of the problem and differ substantially. They give relationships between the film thickness *h* and the capillary number *Ca*. For example, the paper [5] demonstrated theoretically that different physical properties for the different liquids result not in a single function but in a family of the functions *h*(*Ca*).The complete set of previous experimental work fitted the family of curves, while the previous theories could satisfy just some of this experimental data. The solution was found applying the nonlinear theory [5]. The inertial terms and two-dimensional flow together with the parameter of liquid physical properties were accounted. The direct method of Galerkin was applied for solution of the nonlinear problem; therefore the new theory has got an advantage of accurately determining the shape and size of the upper meniscus profile. With the complete set of the available experimental data achieved an excellent agreement with the theoretical results. The classical formula [4] was derived more in detail in [6]. For the case of thick films, it has been given without strict derivation, showing, in particular, that in this case the thickness of a layer does not depend on the surface tension of a liquid.

The classical coating problem of determining the asymptotic film thickness on a flat plate, which is being withdrawn vertically from an infinitely deep liquid pool, has been examined through a numerical solution of the stationary Navier-Stokes equations [7]. For the creeping flow, the dimensionless load *q* was determined as a function of the capillary number *Ca*. For $Ca<0.4$, an agreement of the Wilson's extension [8] with the Levich's well-known expression was found. But for $Ca\to\infty$, *q* asymptotes to 0.582, below the value of 2⁄3 by Deryagin and Levi [9]. For the finite Reynolds numbers $Re\equiv mCa^{3/2}$, where *m* is a dimensionless number involving only the gravitational acceleration *g* and the properties of the fluid, *q* was found independent of the *m* at a given *Ca*. Nevertheless, it was revealed correct only up to a critical capillary number *Ca**, dependent on *m*, beyond which their numerical scheme failed. Similarly, the corresponding nondimensional flow rate $q_\alpha\equiv q(\cos\alpha)^{1/2}$ depends on both *Ca* and *α*for the creeping flows in case of the inclined plate (at an angle *α* to a vertical). Its maximum has been found to increase monotonically with *α* up to 2⁄3 when *α* exceeds a critical angle $\alpha_c\sim\pi/4$, where the plate was inclined midway to the horizontal with its coating surface on the topside. Nonlinear free coating onto a vertical surface was studied theoretically in [10].

When a vessel of liquid has been emptied and put aside, a thin film of liquid clings to the inside and gradually drains down to the bottom under the action of gravity [11]. The layer being thin, the motion is very nearly laminar flow, and the curvature of the surface in a horizontal direction may be ignored. Thus the problem for a cylindrical vessel is reducible to that of a wet plate standing vertically.

Experimental study [12] enhanced the fundamental understanding of the coating processes in a wide range of the varying parameters. They revealed the phenomena of the formation of an asymptotic meniscus profile leading to a development of a cusp at an interface. The dimensionless description of such phenomena allowed identification of the main parameters. And flow visualization revealed the entire flow structure with using a visible laser. The two phenomena of a free coating have been shown depending on the property number *Po*. By parameter *Po* over about 0.5, the dimensionless final film thickness $h_0$is constant up to the capillary number *Ca* of about 1. By *Po* less than 0.1, film thickness $h_0$ depends on *Ca* and the Reynolds number *Re* but it becomes constant when the Weber number $We=Ca\,Re$is less than about 0.2. In both cases $h_0$ is constant when the effect of a surface tension on the meniscus becomes weak. A cusp formation is caused by the inertia (*Re*). By large *Re*, the effect of applicator dimensions on $h_0$ was investigated for flows too.

The thin liquid sheet was transported by a vertical flat plate, which was stationary moving upwards under an action of gravity [13]. Some liquid flowed down, a trend that can be increased by blowing up the air jet on the side layer. The preformed analysis of possible solutions of the stationary flow led to the correlation of the thickness of final layer with the strength of a jet. For the testing of stability, the corresponding non-stationary flows have been investigated, which have shown that the stripped flow is resistant to the long-wave perturbations.

The size and shape of meniscus profiles, which were enlarged by flow, have been measured experimentally [14] and photographically for a range of flow conditions. The free coating of flat sheets withdrawn from a pool of wetting liquids was studied. The withdrawal speeds were varied over several capillary numbers *Ca* below 1 using and oil with viscosity 0.194 Ns/m$^2$ (194 cP). The deformed profiles were modeled by three-parameter analytical expression. The parameters may be used to study influence of *Ca*, coating speed, surface tension, viscosity, density on the profile size and shape. Influence of Re on the profile was noted at Re above 2.

By withdrawn of a body from a liquid bath a liquid film is kept on a surface of the body. In a review [15], after recalling the predictions and results for pure Newtonian liquids coated on the simple solids, an analysis of the deviations to this ideal case was done exploring successively three potential sources of complexity: the liquid-air interface, the bulk rheological properties of the liquid and the mechanical or chemical properties of the solid. For these different complex cases, the significant effects on the film thickness were observed experimentally and summarized the theoretical analysis from the literature.

The propagation of hydrodynamic modes on the surface of agarose gels in the frequency range $10^1$–$10^3$Hz has been studied [16] using the electrically excited surface waves; and rheometry determined the bulk rheological behavior in the frequency range $10^{-2}$–$10^2$Hz. Propagation of the two surface modes, the capillary and the elastic one, was observed at low frequencies, while the regular capillary behavior was detected above a well-defined crossover frequency. The both, theoretical analysis, as well as the measured bulk viscoelastic properties revealed an excellent agreement with the experimental data.

From the recent study of the new soft-micro technologies, the hydrodynamic theory of surface waves propagating on viscoelastic bodies enforced this field of technology with the interesting predictions and the new available applications [17]. Presently many soft small objects, deformable meso- and micro-structures, and macroscopically viscoelastic bodies fabricated from colloids and polymers are produced. Therefore, the new soft products fabricated by functional dynamics based on the mechanical interplay of the viscoelastic material with the medium through their interfaces. In this review, the author recapitulated the field from its birth and theoretical foundation in the latest 1980s up today, through its flourishing in the 90s from the prediction of extraordinary Rayleigh modes in coexistence with ordinary capillary waves on the surface of viscoelastic fluids, a fact first confirmed in experiments with soft gels [16]. With this observational discovery at sight, it was not only settled the theory previously formulated, but mainly opened a new field of applications with soft materials where the mechanical interplay between surface and bulk motions matters. Also, the new unpublished results from surface wave experiments performed with soft colloids were reported in this contribution, where the analytic methods of wave surfing synthesized together with the concept of coexisting capillary-shear modes were claimed as an integrated tool to insightfully scrutinize the bulk rheology of soft solids and viscoelastic fluids.

Considerable work has been done on coating films, summarized in the review articles [18-20]. In particular, work on coating flows includes [12, 21]. They found that the final film thickness depends on the physicochemical properties of the liquid and the withdrawal rate. Pre-metered coating processes attempt to overcome the limitations of free coating. Within certain operating limits, the size of the final film thickness becomes an independent parameter [22-26]. Other experiments on coating flows are available [14, 27, 28]. Based on those studies much is known about the final film thickness and the interfacial profile over a wide range of capillary number. The main interest of work [2] was dip coating at high Reynolds number. Although some investigators analyzed coating flows at high capillary and Reynolds number by approximate methods, e.g. [29, 30], no systematic experimental studies have been performed in the past under those conditions. By the high Reynolds number, the flow in the film and in the coating applicator becomes important. Schweizer [31] has experimentally determined the two-dimensional flow field for a slide-coating device at a maximum capillary number of 0.25. No detailed study of the flow field in dip coating is available. While numerous authors have studied falling thin films, studies by [30] show that rising films are uniquely different.

The submersible coating is to immerse the substrate in a reservoir containing a film forming fluid, and then withdraw from the bath to produce a film. The purpose of [32, 33] was a development of a mathematical model for the hydrodynamic process of immersion, given that the film-forming fluid behaves as a generalized Newtonian fluid. An analytical and simple mathematical model that binds the main parameters of a liquid with the use of the generalized Herschel-Bulkley model was proposed. This model was obtained on the basis of strict balance of mass and momentum applied to the homogeneous on-evaporative system, where the main forces are viscous and gravitational. The parameters that can be evaluated are the velocity profile, flow rate, local thickness, and average thickness of the coating film. Finally, sufficient conditions for the model were obtained. Experimental testing and sensitivity analysis have been presented in the supplementary article as part 2.

## 3. Statement of the problem by nonlinear wave flow on the withdrawn surfaces

It is well known that the thin liquid sheet on the withdrawn surface decreases to the constant thickness $h_0$ determined by the surface moisten quality, its moving velocity $u_0$ and physical properties of fluid: viscosity $\mu$, density $\rho$, surface tension $\sigma$, etc. [34].But earlier investigations did not take into account that the film flow effected by gravitational forces are marked by nonlinearity and has many different regimes including solitary waves [35-39] that strongly influences on the surface covering properties and their quality. The coordinate system $x,y$ shown in Fig. 1 is used.

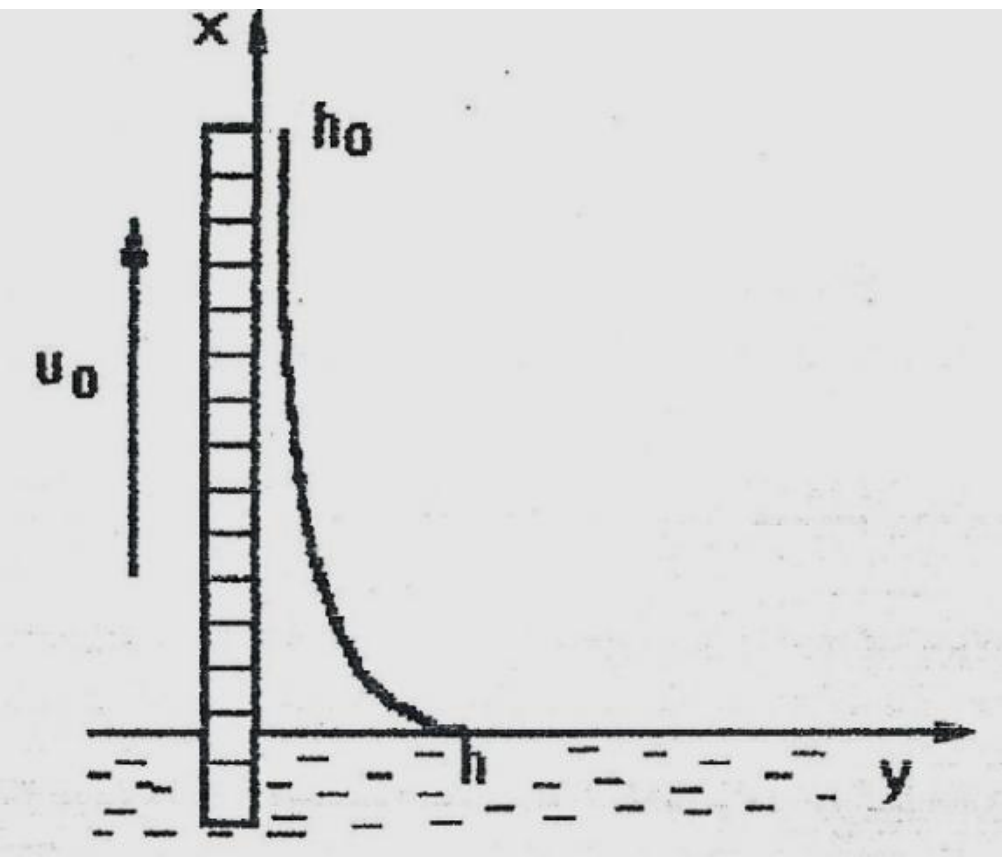

Fig. 1. Model of fluid adhering to a moving withdrawn surface

We consider the problem taking into account influence of the nonlinearity phenomena. It is supposed that the fluid is Newtonian and the process is isothermal. Gravitational force acts against the surface moving direction *x*. It could be also organized using the electromagnetic systems for the process control if the fluid is electroconductive [37-39], e.g. by application of the crossed E, H fields. Despite the above-mentioned works, investigation of the film flow is still interesting being unknown in many aspects because an interplay of the different forces creates a lot of combinations of the diverse regimes. For example, in a considered in this paper flow there are two specific peculiarities: the flow is going against the surface moving direction (vortex flow) and the static meniscus determines the flow beginning part, except the fact that by different Reynolds numbers the situation is changing dramatically.

The substitutive equation array describing the process was derived, assuming that fluid flow is two-dimensional and considering the film flow in a boundary layer approach. The system of the boundary layer equations with the corresponding boundary conditions is

$$\frac{\partial u}{\partial x} = -\frac{\partial v}{\partial y}, \quad \frac{\partial p}{\partial y} = 0\,, \qquad (1)$$

$$\frac{\partial u}{\partial t} + u\frac{\partial u}{\partial x} + v\frac{\partial u}{\partial y} = -\frac{1}{\rho}\frac{\partial p}{\partial x} + \nu\left(\frac{\partial^2 u}{\partial x^2} + \frac{\partial^2 u}{\partial y^2}\right) - g\,, \qquad (2)$$

$$y = 0\,, \quad u = u_0, \qquad v=0; \qquad (3)$$

$$y = \zeta(x,t)\,, \qquad \frac{\partial u}{\partial y} = -\frac{\partial v}{\partial x}, \quad v_\zeta = \frac{\partial \zeta}{\partial t} + u_\zeta \frac{\partial \zeta}{\partial x}\,, \qquad (4)$$

$$p_1 - p_2 = 2\mu\frac{\partial v}{\partial y} - \sigma\frac{\partial^2 \zeta}{\partial x^2}\,, \qquad (5)$$

where $u, v$ are components of the fluid velocity in coordinate system *x,y; p* is the total hydrodynamic pressure, *g*- gravitational acceleration, $y = \zeta(x,t)$- the free surface equation for the film flow, $p_1$- inside pressure value on the free surface, $p_2$- outside pressure value on the free surface, *t*- time, $\nu = \mu/\rho$. Thus, from (1) follows $p=p(x,t)$ – inside pressure of the film flow, $p_2$=const – outside pressure of the film flow (atmospheric).

## 4. Derivation of integral correlations and differential equations for film flow surface

The differential equation array (1), (2) with the boundary conditions (3) - (5) was integrated across the boundary layerusing the Leibniz's rule for differentiation under the integral sign

$$\frac{\partial}{\partial x}\int_0^\zeta u\,dy = \int_0^\zeta \frac{\partial u}{\partial x}dy + u_\zeta \frac{\partial \zeta}{\partial x}\,.$$

For the first equation (1), with account of (3) and the last boundary condition (4) we get

$$\frac{\partial \zeta}{\partial t} + u_\zeta \frac{\partial \zeta}{\partial x} + \frac{\partial q}{\partial x} - u_\zeta \frac{\partial \zeta}{\partial x} = 0, \rightarrow \frac{\partial \zeta}{\partial t} + \frac{\partial q}{\partial x} = 0,$$

where *q* is determined as $q = \int_0^\zeta u\,dy$. Indexes $\zeta$ and 0 indicate here that correspondent values are taken by $y=\zeta$ and $y$=0. Then integrating the equation (2), with account of the boundary conditions (3) - (5) results in

$$\frac{\partial q}{\partial t} - u_\zeta \frac{\partial \zeta}{\partial t} + \frac{\partial}{\partial x}\int_0^\zeta u^2 dy - u_\zeta^2 \frac{\partial \zeta}{\partial x} + \left(vu\right)_\zeta = -\frac{1}{\rho}\frac{\partial p}{\partial x}\zeta - g\zeta + \nu\left(\frac{\partial u}{\partial y}\right)_0^\zeta - \nu\frac{\partial}{\partial x}\left(\frac{\partial \zeta}{\partial t} + u_\zeta \frac{\partial \zeta}{\partial x}\right),$$

$$\left(vu\right)_\zeta = u_\zeta\left(\frac{\partial \zeta}{\partial t} + u_\zeta \frac{\partial \zeta}{\partial x}\right), \quad p = p_a + 2\mu\frac{\partial v}{\partial y} - \sigma\frac{\partial^2 \zeta}{\partial x^2} = p_a - 2\mu\frac{\partial u_\zeta}{\partial x} - \sigma\frac{\partial^2 \zeta}{\partial x^2}\,,$$

where $p_a$ is atmospheric pressure. We used the correlations:

$$v\frac{\partial u}{\partial y}=\frac{\partial(uv)}{\partial y}-u\frac{\partial v}{\partial y}=\frac{\partial(uv)}{\partial y}+\frac{1}{2}\frac{\partial u^2}{\partial x},\quad \left(\frac{\partial v}{\partial x}\right)_0^{\zeta}=\frac{\partial}{\partial x}\left(\frac{\partial \zeta}{\partial t}+u_\zeta\frac{\partial \zeta}{\partial x}\right),$$

$$\left(\frac{\partial u}{\partial y}=-\frac{\partial v}{\partial x}\right)_\zeta,\qquad \frac{\partial u}{\partial x}=-\frac{\partial v}{\partial y},\rightarrow\frac{\partial^2 u}{\partial x^2}=-\frac{\partial}{\partial x}\left(\frac{\partial v}{\partial y}\right)=-\frac{\partial}{\partial y}\left(\frac{\partial v}{\partial x}\right).$$

$(\partial v/\partial x)_0=0$ because $v=0$ on a surface of the plate. The above-considered yields the following equation array for the film flow:

$$\frac{\partial \zeta}{\partial t}+\frac{\partial q}{\partial x}=0,\quad p=p_a-2\mu\frac{\partial u_\zeta}{\partial x}-\sigma\frac{\partial^2\zeta}{\partial x^2},\tag{6}$$

$$\frac{\partial q}{\partial t}+\frac{\partial}{\partial x}\int_0^\zeta u^2dy=-\frac{1}{\rho}\frac{\partial p}{\partial x}\zeta-g\zeta-2\nu\frac{\partial}{\partial x}\left(\frac{\partial \zeta}{\partial t}+u_\zeta\frac{\partial \zeta}{\partial x}\right).$$

Substituting the pressure into the last equation we can get the following equation array of two equations:

$$\frac{\partial \zeta}{\partial t}+\frac{\partial q}{\partial x}=0,\quad \frac{\partial q}{\partial t}+\frac{\partial Q}{\partial x}=\left(2\nu\frac{\partial^2 u_\zeta}{\partial x^2}+\frac{\sigma}{\rho}\frac{\partial^3\zeta}{\partial x^3}\right)\zeta-g\zeta-2\nu\frac{\partial}{\partial x}\left(\frac{\partial \zeta}{\partial t}+u_\zeta\frac{\partial \zeta}{\partial x}\right).\tag{7}$$

Here the function $Q=\int_0^\zeta u^2dy$ was introduced (kinetic energy).

## 5. General case of the uneven surface of the withdrawn plate

Instead of the boundary condition (3) more general condition may be considered for the uneven plate:

$$y=\varepsilon(x),\quad u=u_0,\quad v=0;\tag{8}$$

here $y=\varepsilon(x)$ is equation of the surface of the withdrawn plate (e.g. wavy). Then similar to the above:

$$\frac{\partial}{\partial x}\int_\varepsilon^\zeta udy=\int_\varepsilon^\zeta\frac{\partial u}{\partial x}dy+u_\zeta\frac{\partial \zeta}{\partial x}-u_\varepsilon\frac{\partial \varepsilon}{\partial x}=\int_\varepsilon^\zeta\frac{\partial u}{\partial x}dy+u_\zeta\frac{\partial \zeta}{\partial x}-u_0\frac{\partial \varepsilon}{\partial x},$$

$$\frac{\partial \zeta}{\partial t}+u_\zeta\frac{\partial \zeta}{\partial x}+\frac{\partial q}{\partial x}-u_\zeta\frac{\partial \zeta}{\partial x}+u_\varepsilon\frac{\partial \varepsilon}{\partial x}=0,\rightarrow\frac{\partial \zeta}{\partial t}+\frac{\partial q}{\partial x}+u_0\frac{\partial \varepsilon}{\partial x}=0,$$

$$\frac{\partial \varepsilon}{\partial t}=0,\quad \left(\frac{\partial u}{\partial y}\right)_\varepsilon=0,\quad u_\varepsilon=u_0,$$

$$\frac{\partial q}{\partial t}-u_\zeta\frac{\partial \zeta}{\partial t}+\frac{\partial}{\partial x}\int_\varepsilon^\zeta u^2dy-u_\zeta^2\frac{\partial \zeta}{\partial x}+u_0^2\frac{\partial \varepsilon}{\partial x}+(vu)_\zeta+$$

$$=\frac{1}{\rho}\frac{\partial p}{\partial x}(\varepsilon-\zeta)+g(\varepsilon-\zeta)+\nu\left(\frac{\partial u}{\partial y}\right)_\varepsilon^\zeta-\nu\frac{\partial}{\partial x}\left(\frac{\partial \zeta}{\partial t}+u_\zeta\frac{\partial \zeta}{\partial x}\right).$$

In this general case of uneven surface a the plate, the equation array (7) with the boundary conditions (8) is transformed as follows

$$\frac{\partial \zeta}{\partial t}+\frac{\partial q}{\partial x}+u_0\frac{\partial \varepsilon}{\partial x}=0,\tag{9}$$

$$\frac{\partial q}{\partial t}+\frac{\partial Q}{\partial x}+u_0^2\frac{\partial \varepsilon}{\partial x}=\left(2\nu\frac{\partial^2 u_\zeta}{\partial x^2}+\frac{\sigma}{\rho}\frac{\partial^3\zeta}{\partial x^3}\right)(\zeta-\varepsilon)+g(\varepsilon-\zeta)-2\nu\frac{\partial}{\partial x}\left(\frac{\partial\zeta}{\partial t}+u_\zeta\frac{\partial\zeta}{\partial x}\right).$$

Here $q$ and $Q$ are determined as $q=\int_\varepsilon^\zeta u dy$, $Q=\int_\varepsilon^\zeta u^2 dy$.

## 6. Unique peculiarities of the derived equation array

First of all, as shown above, the similar nonlinear terms in the differential equations obtained after integration across the film layer mutually reduced, so that in the equation arrays (7) and (9) only the second equations contain the nonlinear terms in their right hands.

The other unique feature has concern to the first equations of the system (7) and (9), which are analyzed in the following form:

$$\frac{\partial\zeta}{\partial t}+\frac{\partial q}{\partial x}=0,\quad \frac{\partial\zeta}{\partial t}+\frac{\partial(q+u_0\varepsilon)}{\partial x}=0,$$

where from follows

$$\zeta=q/1=f(x-t\cdot 1),\quad \zeta=(q+u_0\varepsilon)/1=F(x-1\cdot t),\qquad (10)$$

so that the hyperbolic-type equations of the mass conservation in a film flow have general solution $f$ and $F$ – any arbitrary functions of the argument $\eta=x-1\cdot t$, which means that a speed of the waves in both cases is the same 1. In the solutions (10) the 1 is introduced as a unity of velocity (e.g. 1m/s) to correlate the dimensional values. In a dimensionless form it is not needed of course. Thus, $\zeta=q=f(\eta)$, $\zeta=q+u_0\varepsilon=F(\eta)$, $\eta=x-t$.

## 7. Dimensionless equation arrays for film flow on a surface of the plate

The first simple equation (7) shows that with decrease of the flow rate at the current point of $x$ the film thickness has tendency to grow, and inversely with increase of the flow rate – like in the Bernoulli equation: with increase of the width of a flow (tube) the velocity is going down. Here the speed of a plate is constant, therefore with increase of thickness of a film the flow rate may decrease only due to changes in velocity profile. The second equation, except the functions $q$ and $\zeta$ contains the values: velocity of a film flow at the free surface $u_\zeta$, its second derivative by $x$ and to the left – integral from square of $u$.

Therefore, for further analysis of the wave processes in a film flow we need to approximate the film flow profile with a polynomial function: $u=u_0[1+(a_2+a_3y)y^2]$. We obtained this profile using the boundary condition (3) and condition y=0, $\partial u/\partial y=0$ as a requirement of the smooth velocity profile due to the fact that some finite layer of a liquid is kept on a surface of the plate, so that close to a plate in some sublayer velocity is nearly constant. Due to a gravity action against the direction of the plate's movement there is vortex flow in a film because a liquid on a plate is going up, while the free surface is prone to an action of gravity down, up to the point, where gravity becomes small comparing to adhesive, capillary and viscous forces, so that the film flow velocity profile becomes nearly uniform by its thin cross section.

We have done approximation up to a third order by $y$ because it looks reasonable due to vortex flow and change of sign in a layer of a film. For this reason, the parabolic profile seems to be too rough.

The peculiarities of the above equations (6), (7) is that that we do not know the initial conditions, neither boundary at the pool, we only can request stationary parameters far away from the surface of pool:

$$x=\infty,\quad \zeta=h_0,\quad u=u_0,\quad q=q_0=u_0h_0.\qquad (11)$$

As the initial condition we can state capillary meniscus on the plate at the initial moment of time. Equations (6) or (7) with boundary conditions (11) can be used for analysis of the nonlinear film flow that determine the quality of a fluid adhering to the withdrawn surface.

Using the obtained film flow profile, we get the following equations:

$$\frac{\partial \zeta}{\partial t}+\frac{\partial q}{\partial x}=0\,, \qquad (12)$$

$$\frac{\partial}{\partial x}\left\{\frac{\partial \zeta}{\partial t}+u_0\left[1+\zeta^2(a_2+a_3\zeta)\right]\frac{\partial \zeta}{\partial x}\right\}=-u_0(2a_2+3a_3\zeta)\zeta\,, \qquad (13)$$

$$\frac{\partial q}{\partial t}+\frac{\partial Q}{\partial x}=\left\{2\nu u_0\frac{\partial^2}{\partial x^2}\left[\zeta^2(a_2+a_3\zeta)\right]+\frac{\sigma}{\rho}\frac{\partial^3\zeta}{\partial x^3}\right\}\zeta-g\zeta+2\nu u_0(2a_2+3a_3\zeta)\zeta\,. \qquad (14)$$

The equation (13) was obtained from the first boundary condition (4). With the introduced polynomial profile, the equation array could be presented totally through the function $\zeta$ but we use it only for the functions $\partial^2 u_\zeta/\partial x^2$ and $u_\zeta$. After getting the solution, we can substitute into the approximations for calculation of the constants $a_2, a_3$:

$$u=u_0\left[1+(a_2+a_3 y)y^2\right],\quad u_\zeta=u_0\left[1+(a_2+a_3\zeta)\zeta^2\right],$$

$$\frac{\partial u_\zeta}{\partial x}=u_0\frac{\partial}{\partial x}\left[(a_2+a_3\zeta)\zeta^2\right],\ \frac{\partial^2 u_\zeta}{\partial x^2}=u_0\frac{\partial^2}{\partial x^2}\left[(a_2+a_3\zeta)\zeta^2\right],\int_0^\zeta u\,dy=u_0\left(1+\frac{a_2}{3}\zeta^2+\frac{a_3}{4}\zeta^3\right)\zeta\,,$$

$$\int_0^\zeta u^2 dy=u_0^2\left(1+\frac{2}{3}a_2\zeta^2+\frac{a_3}{2}\zeta^3+\frac{a_2^2}{5}\zeta^4+\frac{a_2a_3}{3}\zeta^5+\frac{a_3^2}{7}\zeta^6\right)\zeta\,.$$

The boundary and initial conditions for the equations (12) - (14) are following:

$$x=0,\ \zeta=h_{00},\ q=q_{00},\ Q=Q_{00};\quad x=\infty,\ \zeta=h_*,\ q=q_*,\ Q=Q_*,\frac{\partial \zeta}{\partial x}=\frac{\partial q}{\partial x}=\frac{\partial Q}{\partial x}=0; \qquad (15)$$

$$t=0,\quad \zeta=h_0(x),\quad q=q_0(x),\quad Q=Q_0(x), \qquad (16)$$

where $\zeta=\zeta_0(x)$ is well-known static meniscus equation, and the other parameters are stated according to this and the speed of the plate withdrawn from a pool. The stars assign the stationary parameters far away from the pool.

The equation array (12) - (14) is transformed to the following dimensionless form:

$$\frac{\partial \bar\zeta}{\partial \bar t}+\frac{\partial \bar q}{\partial \bar x}=0\,, \qquad (17)$$

$$\frac{\partial}{\partial \bar x}\left\{\frac{\partial \bar\zeta}{\partial \bar t}+\left[1+\bar\zeta^2(\bar a_2+\bar a_3\bar\zeta)\right]\frac{\partial \bar\zeta}{\partial \bar x}\right\}=-(2\bar a_2+3\bar a_3\bar\zeta)\bar\zeta\,, \qquad (18)$$

$$\frac{\partial \bar q}{\partial \bar t}+\frac{\partial \bar Q}{\partial \bar x}=\frac{2\bar\zeta}{\mathrm{Re}}\frac{\partial^2}{\partial \bar x^2}\left[\bar\zeta^2(\bar a_2+\bar a_3\bar\zeta)\right]+\frac{1}{We}\bar\zeta\frac{\partial^3\bar\zeta}{\partial \bar x^3}-\frac{\bar\zeta}{Fr^2}+\frac{2}{\mathrm{Re}}(2\bar a_2+3\bar a_3\bar\zeta)\bar\zeta\,. \qquad (19)$$

where *R*e and *W*e are the Reynolds and Weber numbers, $Re=u_0 l_0/\nu$, $We=\rho h_{00}u_0^2/\sigma$, $Fr=u_0/\sqrt{gl_0}$ - the Froude number, *We=Ca Re*. The scale values for nondimensional equations (17) - (19) are the following: $\zeta$ , *x*- $l_0$, *u*- $u_0$, *t*- $l_0/u_0$. We introduce $h_{00}$ as characteristic thickness of the film at the beginning of withdrawn, $l_0$ is the characteristic distance by plate where the film flow is established.

## 8. Dimensionless equation arrays for film flow on a surface of the plate

Analysis of the linear equation (17) of the mass conservation shows that $\eta=\bar x-\bar t$ is the complex variable, so that any function $f(\eta)$ satisfies this equation. Obviously, $\eta=\bar{x}-\bar{t}$ is equation of the simple wave moving with constant speed 1 (the same as a plate is moving). Thus, from (17) - (19) we have

$$\bar{\zeta}=\bar{q}=f\left(\bar{x}-\bar{t}\right)=f\left(\eta\right), \tag{20}$$

$$\frac{d^2\bar{\zeta}}{d\eta^2}+\frac{2\bar{a}_2+3\bar{a}_3\bar{\zeta}}{\bar{\zeta}(\bar{a}_2+\bar{a}_3\bar{\zeta})}\left(\frac{d\bar{\zeta}}{d\eta}\right)^2+\frac{2\bar{a}_2+3\bar{a}_3\bar{\zeta}}{\bar{\zeta}(\bar{a}_2+\bar{a}_3\bar{\zeta})}=0, \tag{21}$$

$$\frac{d\bar{Q}}{d\eta}=\frac{\bar{\zeta}}{We}\frac{d^3\bar{\zeta}}{d\eta^3}+\frac{4\bar{\zeta}}{\mathrm{Re}}(\bar{a}_2+3\bar{a}_3\bar{\zeta})\frac{d^2\bar{\zeta}}{d\eta^2}+\frac{d\bar{\zeta}}{d\eta}-\frac{\bar{\zeta}}{Fr^2}+\frac{2}{\mathrm{Re}}(2\bar{a}_2+3\bar{a}_3\bar{\zeta})\bar{\zeta}\,. \tag{22}$$

The solution procedure is a follows. First the non-linear second-order equation (21) is solved, then a solution obtained is substituted into the equation (22), $d\bar{Q}/d\eta$ is computed through the velocity profile and then the equation obtained is used for calculation of the constants $\bar{a}_2$, $\bar{a}_3$.

Film flow to a moving withdrawn surface is not as simple as it is considered in [34] because the nonlinearity of the process is really strong and it cannot be neglected. The mathematical simulation of the process of fluid adhering to a moving withdrawn surface in linear approach is rough enough that explains poor correspondence between linear theory results and practice one. This process might be also controlled by electromagnetic fields [37-39] in case of electroconductive fluid.

## 9. Concluding remarks

We have identical profiles (20) for the functions of the film flow surface and the film flow rate. The nonlinear second-order equation (21) can be solved for a range of available values of the constants and then from (22), after substitution of the solution obtained, the constants of the polynomial approximation of the profile, which satisfy the (22) can be computed. The (21) shows that parabolic film flow profile ($a_3$=0) leads to a universal solution, which does not depend on the film flow profile. But the constants of integration and constant $a_2$ in a film flow profile can be computed afterwards from the equation (22). Further analysis of the equations derived for a smooth withdrawn plate, as well as for a wavy plate, and their solution is a subject for further research.